\documentclass[12pt]{article}

\usepackage{scicite}

\usepackage{times}

\usepackage{graphicx}

\newenvironment{sciabstract}{%
\begin{quote} \bf}
{\end{quote}}

\newcounter{lastnote}

\title{Absorption Features in the X-ray Spectrum\\ of an Ordinary Radio Pulsar}

\author
{Oleg Kargaltsev$^{1,2\ast}$,  Martin Durant$^{1}$,   Zdenka Misanovic$^3$, and George G. Pavlov$^{4,5}$\\
\\
\normalsize{$^{1}$Department of Astronomy, 222 Bryant Space Science Center,}\\
\normalsize{University of Florida, Gainesville, Florida, 32611 USA}\\
\normalsize{$^{2}$Department of Physics,}\\
\normalsize{George Washington University, Washington, DC, 20052, USA}\\
\normalsize{$^{3}$School of Physics,}\\
\normalsize{Monash University, Melbourne, 3800 VIC, Australia}\\
\normalsize{$^{4}$Department of Astronomy and Astrophysics, 525 Davey Laboratory,}\\
\normalsize{The Pennsylvania State University, University Park, Pennsylvania, 16802 USA}\\
\normalsize{$^{5}$St.-Petersburg State Polytechnic University, Polytekhnicheskaya ul.\ 29, St.\ Petersburg, 195251 Russia}\\
\\
\normalsize{$^\ast$To whom correspondence should be addressed; E-mail:  kargaltsev@gwu.edu }
}

\date{}

\begin{document} 


\baselineskip24pt


\maketitle


\begin{sciabstract}
The vast majority of known non-accreting neutron stars (NSs) are rotation-powered radio and/or $\gamma$-ray pulsars.  So far, their multiwavelength spectra have all been  described satisfactorily  by  thermal and non-thermal  continuum models, with no spectral lines. Spectral features have, however, been  found  in a handful of exotic NSs and thought to be a manifestation of their unique traits. Here we report  the  detection of absorption features  in the  X-ray spectrum of an ordinary rotation-powered radio pulsar, J1740+1000. Our findings bridge the gap between the spectra of pulsars and other, more exotic, NSs, suggesting that the features are more common in the NS spectra than they have been thought so far.
\end{sciabstract}

The study of neutron stars (NSs) offers a unique opportunity to understand physical
processes under the most extreme conditions in the Universe. These super-dense, compact  stars  have extremely strong magnetic fields reaching  up to a few hundred Teragauss. Thanks to the strength   of the magnetic field and rapid rotation, young NSs    are capable of generating and maintaining huge  electric potential drops (up to a few Petavolts) across parts of their     magnetospheres, leading to  efficient particle acceleration, pair cascades and high-energy radiation. The vast majority ($>98\%$) of known     NSs are   rotation-powered   radio or $\gamma$-ray pulsars\cite{2010PNAS..107.7147K} (Fig. 1).   The continuum emission of these objects (both thermal emission from the NS surface and non-thermal from the magnetosphere)  has been extensively studied and has provided    important information about the structure and  properties of pulsar magnetospheres, the composition and temperature of the surface layers, and the state of the NS interior \cite{2006RPPh...69.2631H}.
    
    Recently, absorption features have been detected in the thermal X-ray spectra of a few unusual NSs --  `X-ray dim Isolated NSs' (XDINSs),  which do not emit radio pulsations and likely have empty (or particle-starved) magnetospheres,   Central Compact Objects (CCOs) in supernova remnants, which are radio-quiet NSs with low magnetic fields, and   `Rotating Radio Transients' (RRATs), which belong to a recently discovered class of  NSs  seen in the radio only intermittently.    
 All of these NSs are atypical compared to ordinary, rotation-powered pulsars, and the spectral features have been attributed to peculiarities of these objects and exotic physics. For example, in XDINSs the surface magnetic field is strong enough for QED effects to be important, while for the very weakly magnetized CCOs the electron cyclotron energy for matter at the surface is low enough that it falls into the soft X-ray band. For ordinary pulsars (with magnetic fields $\sim 10^{12}$ G)  neither of these conditions is expected to be satisfied at the NS surface and hence no analogous spectral features are expected in soft X-rays.

The 100\,kyr old radio pulsar J1740+1000   was detected  in 2002\cite{2002ApJ...564..333M}. The pulsar's spin period, $P=154$ ms, and its derivative, $\dot{P}=2.15\times10^{-14}$\,s\,s$^{-1}$, place it among other relatively young pulsars in 
 the  $P$-$\dot{P}$ diagram (Fig.\ 1), suggesting  that it is a fairly unremarkable  pulsar, which loses rotational energy at a rate $\dot{E}=2.3\times10^{35}$ erg s$^{-1}$ and has a surface magnetic field $B_{0}=1.8\times10^{12}$ G.   
With the dispersion measure DM=24 pc cm$^{-3}$ and Galactic latitude of
$20.3^\circ$,  the derived distance of PSR J1740+1000 is ~1.4\,kpc or 1.2\,kpc
(according to the Galactic electron density models of \cite{1993ApJ...411..674T} and \cite{2002astro.ph..7156C}, respectively).    The off-plane location leads to fortuitously  low interstellar absorption, making  
 this pulsar one of the few whose spectra can be studied in  softer X-rays ($<1$ keV).
  For   10\% ISM ionization, the measured DM  implies a hydrogen column density $N_{\rm H}\approx7.4\times10^{20}$\,cm$^{-2}$. The total Galactic HI column in this direction is $\approx(8.2-8.7)\times10^{20}$\,cm$^{-2}$ \cite{1990ARA&A..28..215D,2005A&A...440..775K }.

   X-ray emission from PSR J1740+1000 was first detected in a 20 ks {\sl Chandra} ACIS observation \cite{2008ApJ...684..542K}.
Here we report the results of   observations  with  the  {\sl XMM-Newton} X-ray observatory European Photon Imaging Camera (EPIC) and  {\sl Chandra X-ray Observatory} Advanced CCD Imaging Spectrograph (see the supplementary text for observation details).  

The phase-integrated spectrum of J1740+1000 (Fig.~2, left panel), measured with  EPIC and ACIS, is well-fit (reduced $\chi_{\nu}^2$=0.92, for $\nu=$81 d.o.f.) in 0.2--10\,keV by a model containing two blackbody (BB) components \cite{note3} (with 
 $T_1^\infty=71^{+5}_{-9}$\,eV,  $T_2^\infty=148^{+16}_{-15}$\,eV  and $R_1^\infty=7.6^{+4.1}_{-2.4}$ km, $R_2^\infty=0.64_{-0.17}^{+0.27}$ km, for $d=1.4$ kpc) plus a power-law (PL) component with photon index $\Gamma=1.6\pm0.6$  and $N_{\rm H}=9.7_{-1.3}^{+1.5}\times10^{20}$\,cm$^{-2}$  \cite{note2}  (here and below uncertainties are given at 68\% confidence for a single parameter).
  The isotropic bolometric luminosity of dominant cooler thermal emission from most of the NS surface is $2\times10^{32}$ erg s$^{-1}$.   The  hotter thermal component with a factor of 140 smaller emitting area could be attributed to a hot spot on the NS surface.  Being faint and hard, the PL component contributes significantly only above $1.3$ keV (Fig. 2). To simplify the analysis of the thermal emission, we only considered energies $<$1.1\,keV and therefore ignore the PL component.  The emission  shows substantial pulsations  with the pulsed fraction of $30\%\pm4\%$ in 0.2--1.1 keV (Fig.~3, left). Overall, the properties  resemble those of other nearby middle-aged pulsars  (Geminga, B0656+14, and B1055$-$52\cite{2002nsps.conf..273P,2005ApJ...623.1051D}). 
  
Figure 3 (right) shows  spectra corresponding to three selected phase intervals:  [0.0-0.3] (``dip''),  [0.3-0.6,0.8-1.0]  (``rise and fall'') and [0.6-0.8] (``peak''). Large systematic differences appear between the three spectra. While overlapping  at 0.3--0.5 keV and above 1 keV, the spectra differ markedly in 0.5--0.8 keV and below 0.3  keV. The relative  differences in phase-resolved spectra  cannot result from  poor calibration (as all photons are detected with the same instrument during the same observation), and the residuals strongly suggest two phase-dependent spectral features. Indeed, when we added  a   Gaussian absorption line (GABS) to a single BB  continuum  (BB$\times$GABS, where GABS=$\exp(-\tau e^{-0.5[(E-E_c)/\sigma]^2})$; see Table 1 caption),  the differences between the phase-resolved spectra in the 0.5-0.8 keV range could be accounted for solely by variations in the GABS parameters  with the BB component held fixed. The   0.5-0.8 keV line-like feature  is strong and present in all three phase intervals, but both the line's central  energy  and strength vary with phase, 
becoming progressively stronger  from the pulse maximum to the pulse minimum (Fig.~3). 
  In the BB$\times$GABS fits there are no systematic residuals in the  0.5--0.8 keV range.  The residuals below 0.3 keV vary in concert with the 0.5--0.8 keV absorption line strengths; however, with the relatively small number of photons collected we could not reliably characterize  the parameters of  the low-energy absorption feature which  is at the edge of the EPIC sensitivity range.
 
As an alternative to the absorption feature scenario, we  considered a pure continuum, BB+BB, model.  We conclude that  the absorption feature interpretation is  strongly preferred over the hot-spot scenario (see supplementary material). 

 The more significant of the two spectral features has the central energy, $E=550$--$650$ eV (see Table 1, bottom section), close to the crossover energy for the hot and cold BB components in the pure continuum model. Therefore, because even a two-component model is likely to be an oversimplification (one can expect the NS surface temperature to be a smooth function of the magnetic co-lattitude),  one could speculate that the remaining residuals could be  reduced  by increasing the level of complexity and introducing yet another BB component. However,  this would still not  account for the phase-dependent changes below $\approx0.3$ keV. One could introduce yet another even cooler BB, but   this emission  would have to come from the bulk of the NS surface and hence should not depend on the rotation phase.     Therefore, with the data in hand, the Occam's razor  strongly favors the spectral feature interpretation.

So far, absorption features  were seen only in the spectra of atypical, exotic isolated NSs. These features are expected to provide an important diagnostic of the NS equation of state, atmospheric composition and magnetic field strength\cite{2011ApJ...736..117K,2012ApJ...751...15S}. Various  interpretations have been put forward. Among widely discussed options are  (a) atomic transitions in the magnetized NS atmospheres or condensed surface layers\cite{2007MNRAS.377..905M,2005ApJ...628..902V} and  (b) cyclotron absorption features produced either by electrons in the NS magnetosphere\cite{2002ApJ...574L..61S,2012ApJ...751...15S} or protons (or light ions) in a hot ionized layer near the NS surface\cite{2004ApJ...608..432V}.

Can the explanations used for the absorption features of exotic NSs be applied to the case of J1740+1000? The  energy of the feature we fitted is $\approx0.5-0.7$ keV, while the second, less significant feature is at $0.1-0.25$ keV. These energies are too large for transitions in  a magnetized  hydrogen atom in the $B\approx1.8\times10^{12}$ G surface field\cite{1994asmf.book.....R}.  The energies possibly could match a transition to  an excited state in hydrogen-like ions\cite{2006RPPh...69.2631H}
   for $Z>2$. In this case, the detected spectral features would provide a direct probe of the composition and physical 
conditions of the surface layers (see e.g., \cite{2011A&A...534A..74H} and references therein).

Alternatively, the two features could be the fundamental and the first harmonic of the electron
cyclotron energy $E_{\rm ce}=11.6(B/10^{12}\,\rm{G})$\,keV. However, in this case  the  absorbing/scattering particles must be located in the magnetosphere\cite{1997ApJ...491..296R}, at several stellar radii above the NS surface in a much weaker magnetic field, $B(r)=B_{\rm eq}(1+3\cos^2\theta_B)^{1/2}(R/r)^3$, where $R$ is the NS radius and $\theta_B$ is the magnetic co-latitude. 
  Such a screen would then be analogous to the Earth's van Allen belts.  Such a scenario can also explain the phase dependence of the feature's width and central energy, because the strength of the magnetic field and the density of the absorbing (scattering) screen should depend on the angle between the magnetic dipole axis and the observer's line of sight.  If the spectral features are indeed due to the cyclotron absorption in the magnetosphere,
at least some regions of it must contain nonrelativistic (or mildly relativistic) electrons/positrons. The observed width of the 0.5-0.8 keV feature suggests that the mean velocity of these particles should not exceed 0.2--0.3 c.  It is possible that a   population of ``warm'' electrons exists in the closed line zone forming radiation belts (see \cite{1997ApJ...491..296R}  and \cite{2007MNRAS.378.1481L}).  Because the orientations of the pulsar spin and magnetic axis
with respect  to the observer's line of sight are not known, it is possible that the line of sight  traverses some volume in the closed field line region of the magnetosphere at all rotation phases, i.e.,  
some absorption is always present.  For the spin-down parameters of 
J1740+1000 equation 4  of  \cite{1997ApJ...491..296R}  predicts optical depth of   $\sim 3\times10^{-3}$ at $E=600$ eV, assuming the Goldreich-Julian charge density. However, it is widely accepted that the actual charge density in the pulsar magnetosphere must be larger due to  pair production, and  the  multiplication coefficient of 200--300, required in our case,
 is comfortably below the typically discussed range of  $10^3$--$10^6$ (e.g., \cite{2011MNRAS.410..381B} and references therein). Because it is not clear why heavy ($Z>2$) elements would be exposed only in certain parts on the NS surface (thus leading to phase-dependence), we currently prefer 
 a magnetospheric interpretation of the absorption features.
  
There are three other middle-aged pulsars   (Geminga, B0656+14, and B1055--52) which are similarly nearby, bright, and  have  low  interstellar absorption. The  spectra of these pulsars show  significant variability with  pulse phase, which has been commonly  attributed to the changing view of the non-uniformly heated NS surface\cite{2005ApJ...623.1051D}. However,  even with allowance for such variability, noticeable phase-dependent  residuals\cite{2005ApJ...633.1114J}  in the spectral  fits  are yet to be explained.
The spectral features found for J1740+1000 suggest that the systematic residuals for other pulsars may be better explained by a model with absorption feature(s). Indeed, there is  strong evidence that the lines seen in XDINSs and/or CCOs  also  show strong phase dependence \cite{2004A&A...419.1077H,2009ApJ...695L..35G}.  
The dependence of absorption feature properties on pulsar characteristics can tell us about the physical origin of the features. For example, in the  cyclotron scenario, the energies of the lines  are expected to correlate with the magnetic field strength and possibly with the size of the magnetosphere. Atomic  features would depend upon the pulsar surface density, temperature and composition (as is seen in, e.g., white dwarfs).



\clearpage

{\bf Acknowledgments:} We thank Igor Volkov for useful discussions and advice and Zaven Arzoumanian for his help in preparing the observing proposal. This work was supported
by National Aeronautics
Space Administration grants NNX06AH67G and NNX09AC84G, and through Chandra
Award Number G0-11096A issued by the Chandra X-ray Observatory
Center, which is operated by the Smithsonian Astrophysical
Observatory for and on behalf of the National Aeronautics
Space Administration under contract NAS8-03060.
  The work was also partly supported by the Ministry
of Education and Science of the Russian Federation
(contract 11.G34.310001). The data used in this publication are freely available from  the Chandra Data Archive (http://cxc.harvard.edu/cda/) and the XMM-Newton Science Archive  (http://xmm.esac.esa.int/xsa/).

\clearpage

\begin{table}\small{
\begin{tabular}{c|ccc}
\hline
Parameter  & \multicolumn{3}{c}{Phase interval} \\
  &Dip & Rise and Fall & Peak \\
  &$[0.0-0.3]$ &  [0.3$-$0.6],[0.8$-$1.0] & [0.6$-$8.0]\\
  \hline
&\multicolumn{3}{c}{BB+BB (fixed $T_{1,2}=71,148$\,eV), $\chi^2=64.4$ for 65 d.o.f.} \\
$R^\infty_{\rm 1}$, km
				& $7.4\pm0.2$  & $7.6\pm0.2$ & $8.4\pm0.3$ 
				 \\ 
$R^\infty_{\rm 2}$, km
				& $0.61_{-0.04}^{+0.03}$ & $0.69_{-0.02}^{+0.03}$ & $0.80_{-0.04}^{+0.05}$ 
				 \\ 
\hline
&\multicolumn{3}{c}{BB+BB ($T_{1}$, $T_{2}$ are fitted), $\chi^2=53.7$ for 59 d.o.f.} \\
$T^\infty_{\rm 1}$, eV&  $66\pm 6$&  $67\pm 11$&  $34_{-9}^{+10}$
\\
$R^\infty_{\rm 1}$, km 
                &  $9.2^{+3.1}_{-2.0}$ &  $8.4_{-2.4}^{+5.1}$&  $102_{-72}^{+455}$
                 \\
$T^\infty_{\rm 2}$, eV&  $184_{-11}^{+20}$&  $130_{-14}^{+18}$&  $103_{-4}^{+6}$
\\
$R^\infty_{\rm 2}$, km
                &  $0.35_{-0.10}^{+0.14}$ &  $1.04_{-0.33}^{+0.45}$&  $3.0_{-0.5}^{+0.4}$
                 \\
\hline
&\multicolumn{3}{c}{GABS$\times$BB model, $\chi^2=46.2$ for 60 d.o.f.} \\
$T^\infty_{\rm 1}$, eV   & \multicolumn{3}{c}{$94\pm 2$   (tied)}
\\ 
$R^\infty_{\rm 1}$, km
				& \multicolumn{3}{c}{$4.3\pm0.1$  (tied)}
				 \\ 
 $E_c$, eV   & $646_{-12}^{+13}$  & $635_{-24}^{+11}$  & $548\pm12$  
\\ 
$\sigma$, eV
				& $137_{-13}^{+17}$ & $161_{-14}^{+22}$ & $35_{-15}^{+22}$ 
				 \\ 
$\tau$	 &  $0.92\pm0.14$ & $0.58\pm0.10$\ & $0.80\pm0.57$ \\			 
EW, eV
				& $234\pm34$& $192\pm24$& $54\pm20$\\ 				
\hline
\end{tabular}
\caption{ 
Parameters (with 1$\sigma$ statistical uncertainties) of the spectral models fitted to the phase-resolved spectra in 0.2-1.1 keV range.  In all cases the spectra from three phase  bins have been fitted simultaneously for consistency.  $T^\infty$ (eV) and $R^\infty$ (km)  are BB temperature and radius (as measured at the infinity), respectively. The radii are given for 1.4 kps distance.
For the line, $E_c$, $\sigma$, $\tau$, and EW  are the central energy, width, optical depth and the corresponding equivalent width, respectively. The column density $N_{\rm H}$ is kept constant in every case, at the value of $9.7\times10^{20}$\,cm$^{-2}$ obtained for the phase-integrated spectrum (see text).  EW is not an additional parameter, but is inferred from the other GABS parameters.}
}
\end{table}

\clearpage

\begin{figure}
\includegraphics[width=0.99\textwidth,angle=0]{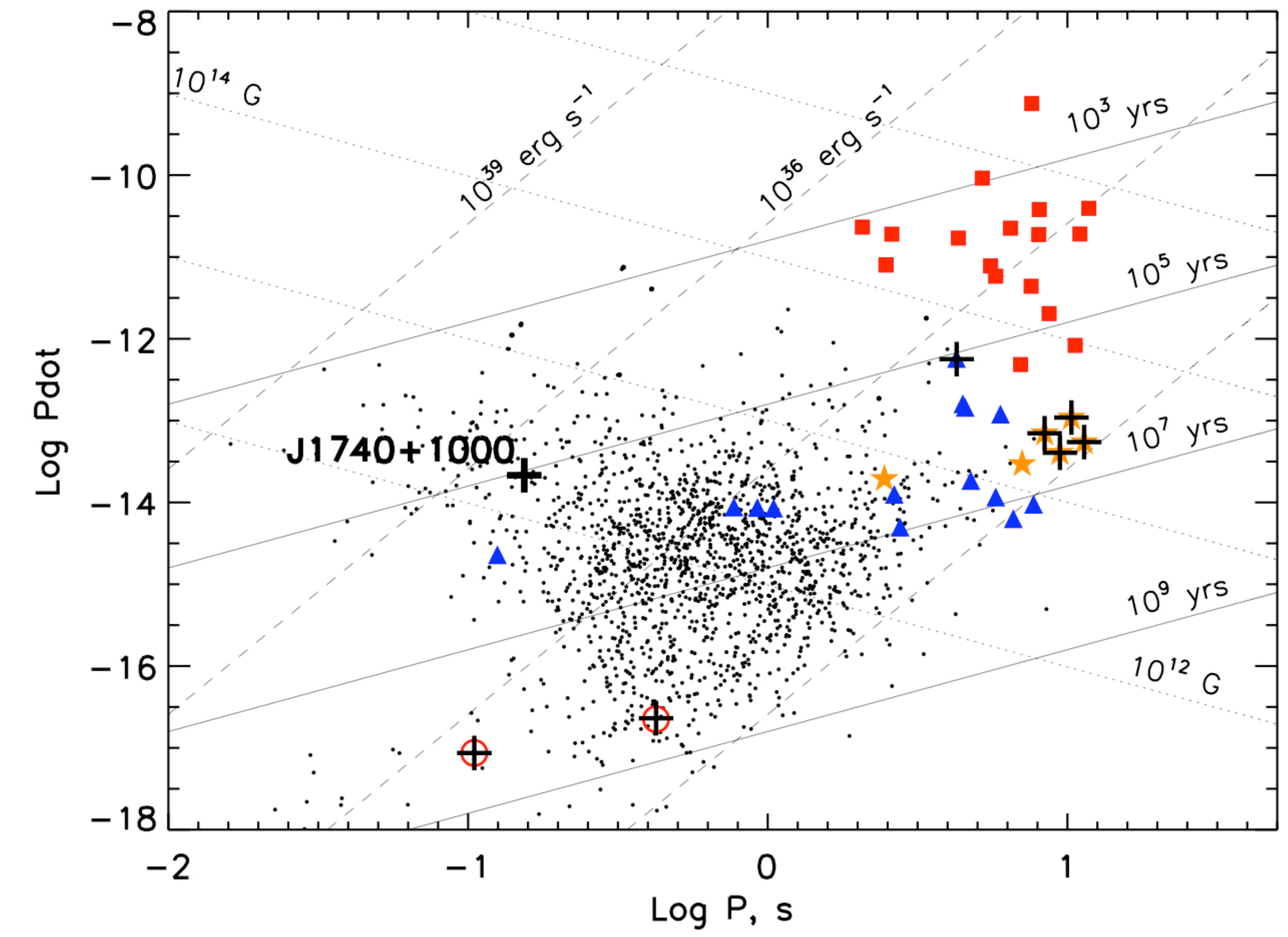}
\caption{ Period $P$ versus  period derivative $\dot{P}$ for  pulsars of various types.
Two parameters are most commonly used to characterize pulsars: the pulsation period $P$, and its rate of change $\dot{P}$. The data come from the ATNF catalog\cite{2005AJ....129.1993M}. A number of physical indicators can be derived from $P$ and $\dot{P}$, namely the spin-down power,  magnetic field, and spin-down   age (over-plotted as labelled straight lines). Whereas the vast majority of ordinary, rotation-powered pulsars (black dots) are grouped in the middle of the Figure, exotic objects such as magnetars (red squares), RRATs (blue triangles), CCOs (red circles,) and XDINSs (orange stars) appear located away from the ordinary pulsar population.  We have marked the NSs with reported spectral features (for which both $P$ and $\dot{P}$ have been determined) with large black crosses. }
\end{figure}

\begin{figure}
\begin{center}\vspace{-0.9cm}
\includegraphics[width=0.78\textwidth,angle=0]{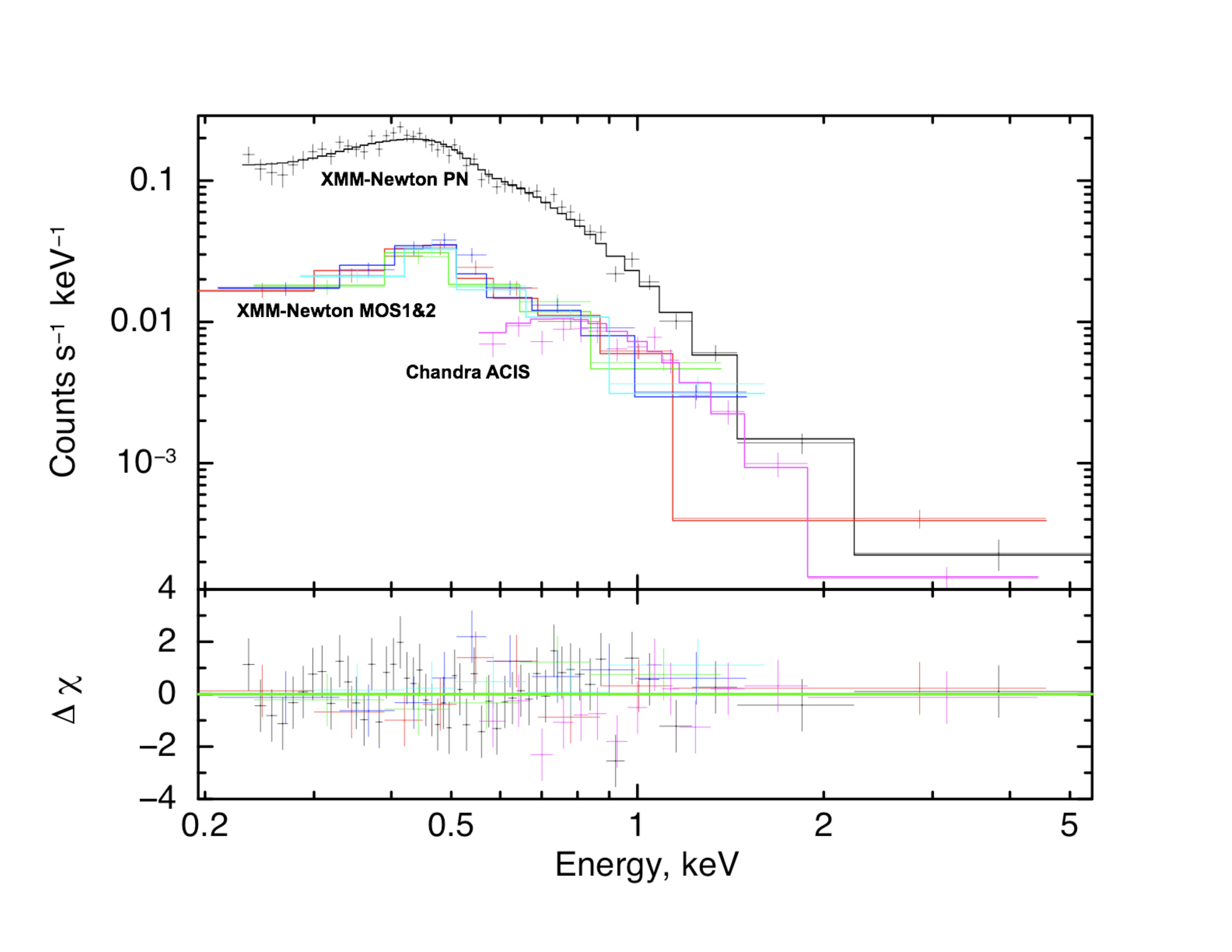}\vspace{-0.8cm}
\includegraphics[width=0.78\textwidth,angle=0]{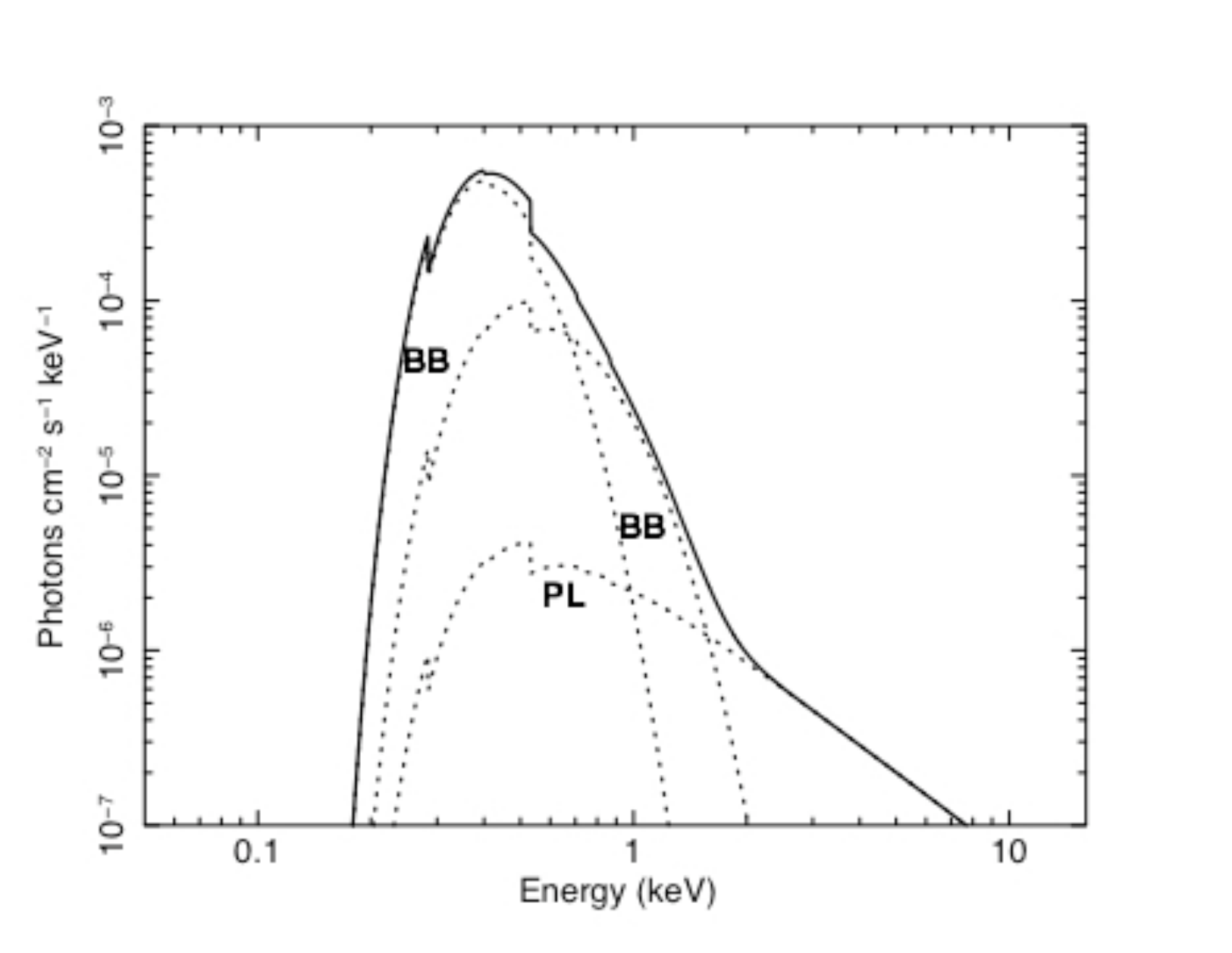}
\end{center}\vspace{-1.0cm}
\caption{ Phase-integrated spectra of PSR J1740+1000.  {\em Left:}  The count-rate spectra obtained with {\sl XMM-Newton} EPIC and {\sl Chandra} ACIS are shown together with the best-fit  absorbed BB+BB+PL model.  The errorbars shown in this and other figures are $1\sigma$ for Poisson statistics. Different colors correspond to different instruments  of  the two X-ray observatories (as labeled). Here and in other spectral plots, the errorbars correspond to the data, and the lines correspond to the best-fit models.  {\em Right:} The best-fit model photon spectrum, with the individual components shown (dotted lines). See text for details of the fitting and parameter values. The jumps in the spectrum  are the edges due to the interstellar  absorption. }
\end{figure}

\begin{figure}
\includegraphics[width=0.99\textwidth,angle=0]{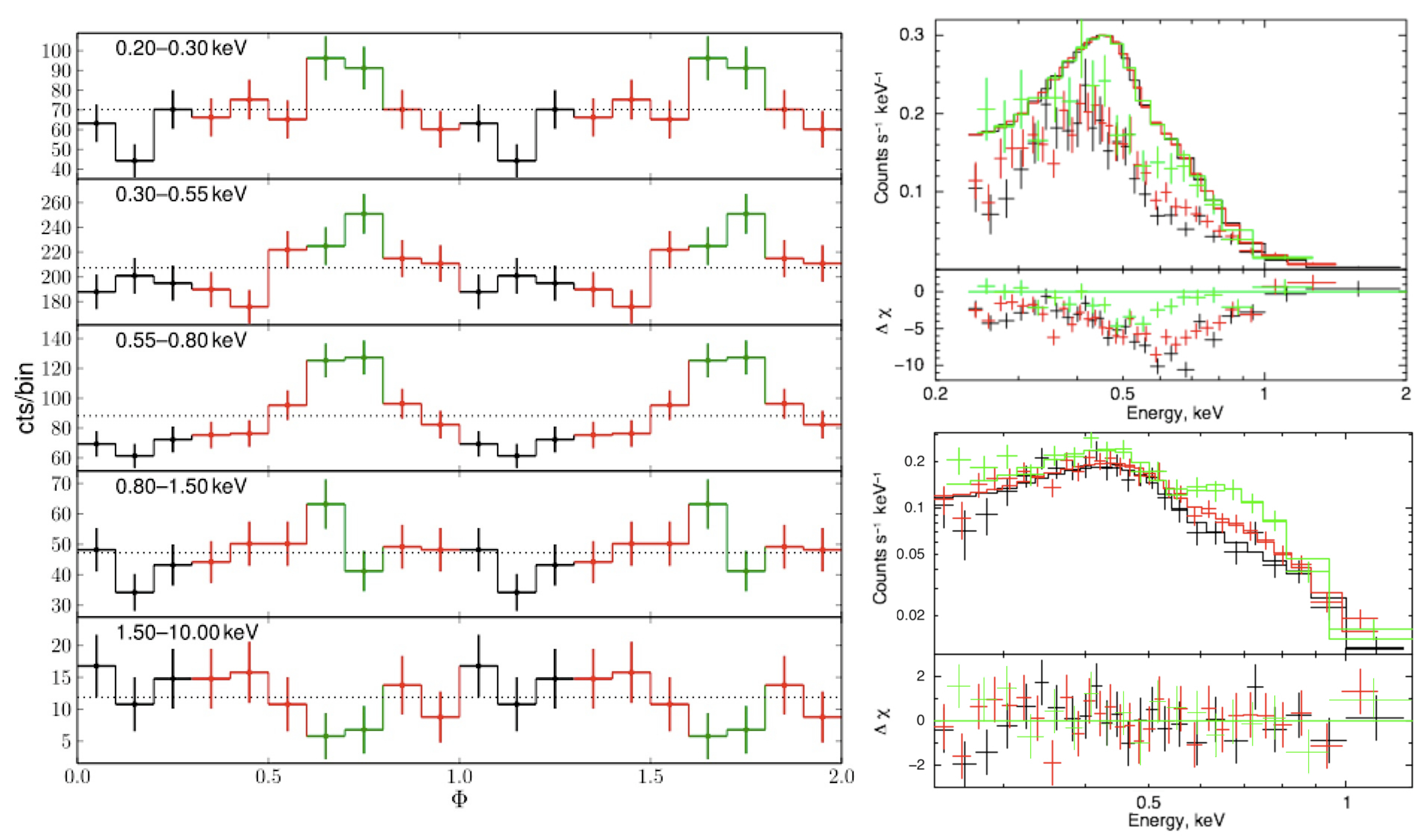}
\caption{  Pulse profiles and phase-resolved spectra of PSR J1740+1000.
 {\em Left:} Energy-resolved background-subtracted pulse profiles measured with 
{\sl XMM-Newton} EPIC PN in five energy bands.
 {\em Top right:} Count-rate spectra constructed by  selecting  the photons from  the three chosen phase intervals 
 shown  by red, green, and black in the left panel. We also show residuals with respect to the BB component (solid lines) of the best-fit  BB$\times$GABS model to emphasize the shape and locations of the lines we observed. 
 {\em Bottom right:}  BB$\times$GABS fits and residuals for the  three chosen phase intervals. No systematic residuals are seen between 0.5 and 0.8\,keV. Here we have not fitted for the low-energy line, leaving systematic residuals below  0.3\,keV.  See supplementary material for details.  }
\end{figure}

 \begin{figure}[b!]
 \vspace{-2.7cm}
 \centering
\includegraphics[width=0.65\textwidth,angle=0]{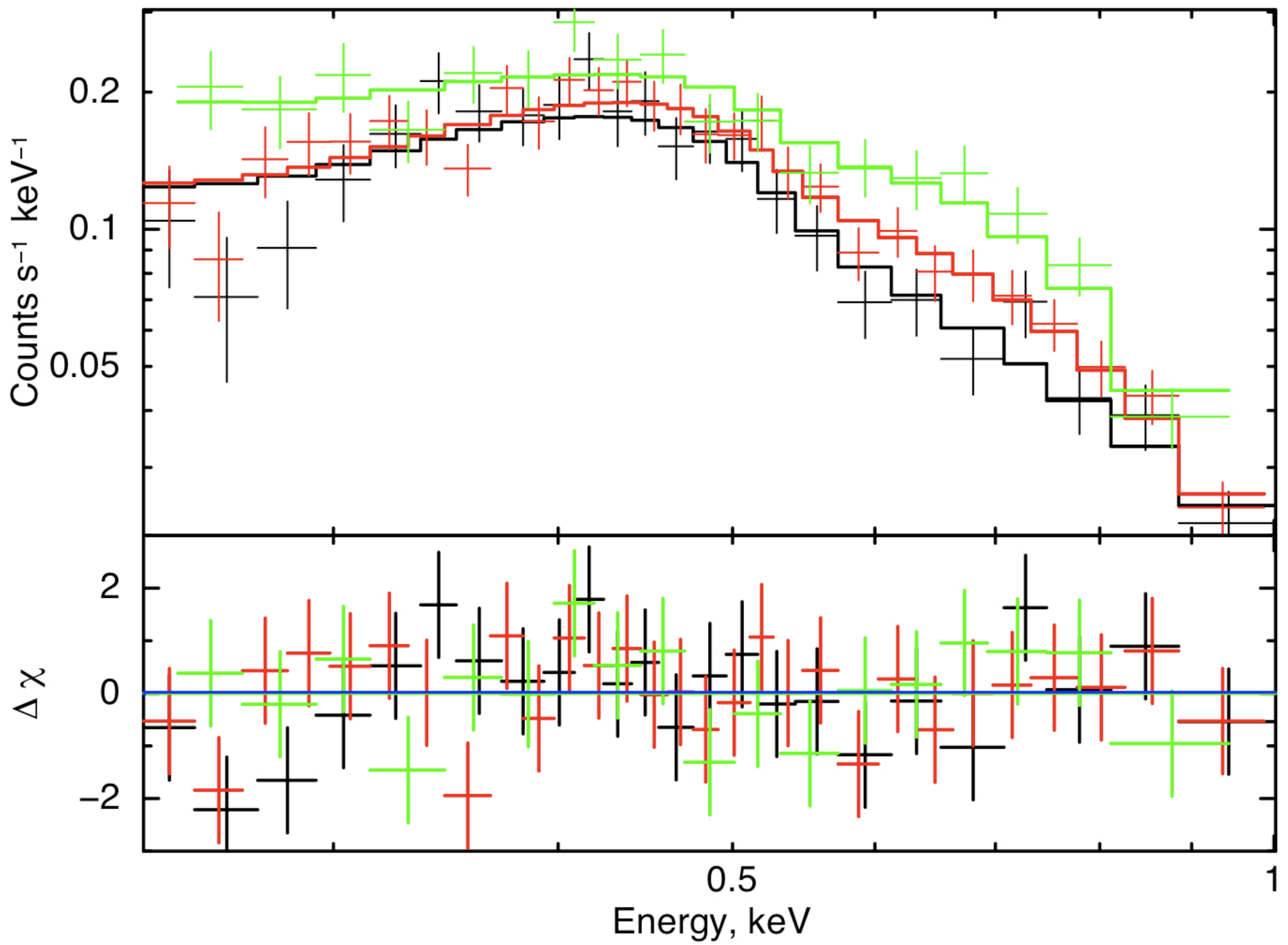}
\includegraphics[width=0.65\textwidth,angle=0]{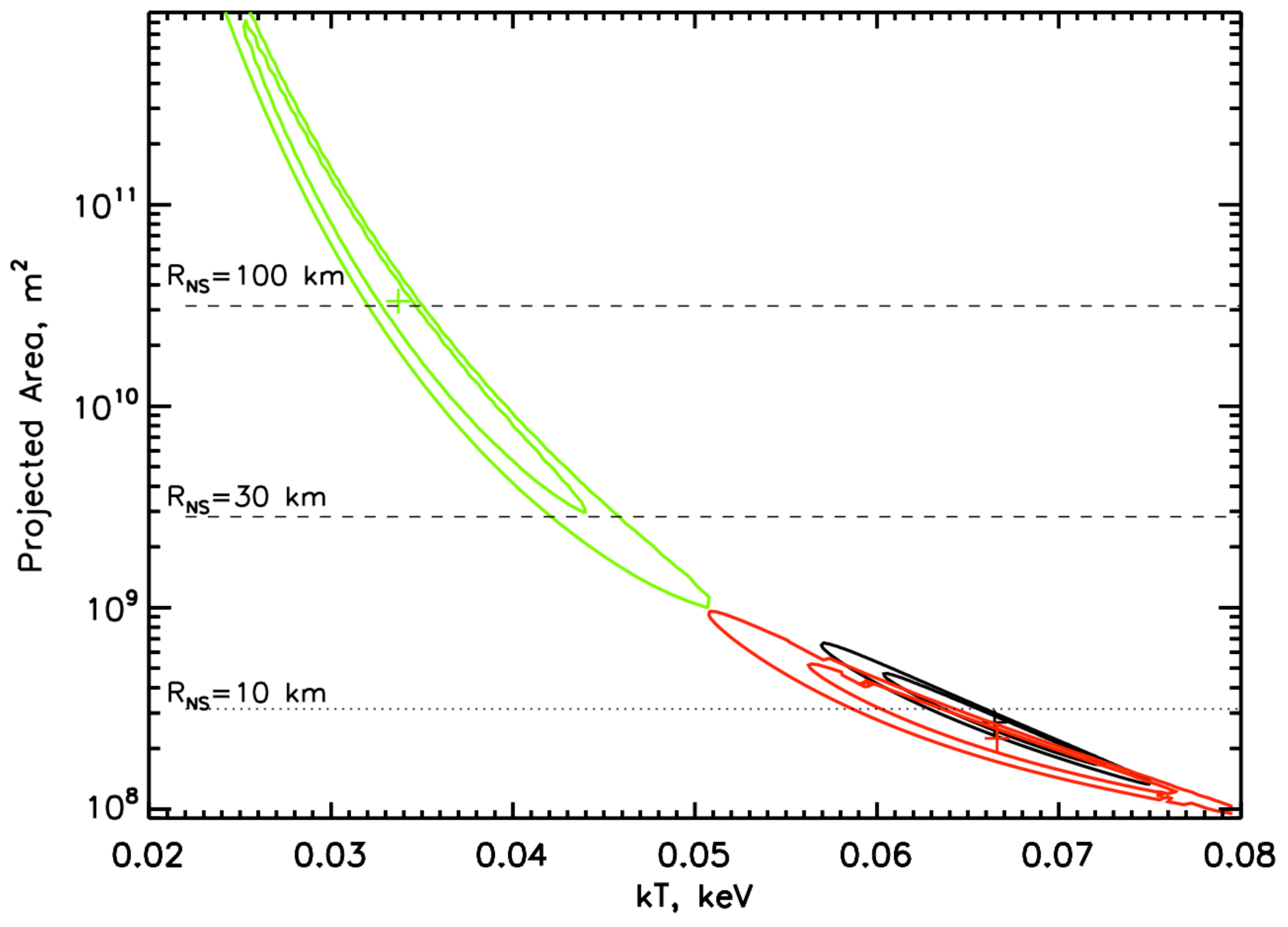}
\caption*{ Figure S1:      {\em Top:} EPIC-PN  count-rate  spectra below 1.3 keV together  with the best-fit BB+BB model for the three chosen phase intervals, leaving  significant   non-zero residuals below 0.3 keV and in   0.5--0.8  keV range. {\em Bottom:}  Confidence contours (68\% and 90\% for single interesting parameter; $\Delta\chi^2$=1.0, and 2.7, respectively) for the cool BB component of the BB+BB fit to the phase 
resolved spectra. During the contour calculation, $N_H$ was frozen at the best-fit value of $9.7\times10^{20}$\,cm$^{-2}$, while the rest of the parameters were re-fitted at each point of the grid.
 }
\end{figure}

\clearpage

\noindent {\bf Supplementary Material}

{\bf Details about X-ray observations:} The EPIC PN detector was operated in Small Window mode, providing a time resolution of 5.7\,ms,  while the MOS1 and MOS2 detectors were in Full Frame  mode with a 2.6\,s time resolution. After correcting for dead time, the useful PN exposure was 46\,ks. The aim of this observation was to analyse the X-ray spectrum as a function of pulse phase (i.e., phase-resolved spectroscopy).  We also acquired a 64.6\,ks  {\sl Chandra} ACIS observation,   with a  much higher angular resolution of $\simeq0.5''$ but a coarser  time resolution of 3.2\,s.   Prior to fitting, all the spectra were grouped to have at least 30 counts per bin. The background contribution is completely negligible for the ACIS data while for the EPIC data it was  8\%--13\% (in 0.2-10 keV), becoming  significant  only  at photon energies $E>5$ keV.  In total we have analyzed 4,448  photons in 0.2--10 keV for EPIC and  397 photons in 0.5--8 keV  for ACIS.  Spectral modeling and fitting was done with XSPEC  v.\ 12.7.0  \cite{1996ASPC..101...17A}.

{\bf Producing pulse profile and phase-resolved spectra:} Pulsars are NSs characterised by their very regular periodic variations in brightness.  PSR J1740+1000 has a period $P=154$ ms (or, equivalently, $\nu=1/P=6.49$ Hz) which generally decreases with time.  
 However, one can directly   measure the period at the epoch of observation from the obtained data, given sufficiently strong pulsed signal. We  used the well-known  $Z_{n}^2$  (Rayleigh)  test   
  to search for the period in  the vicinity  of 154\,ms (more precisely in  6.48-6.50 Hz range) in the EPIC PN data.  The results show a clear and highly significant maximum ($Z_1^2=45.9$) at   $\nu=6.489649(2)$ Hz, which we used to calculate the phase $\phi=\nu t$ and fold  the pulse profiles shown in Figure 3 (left panel) and to extract the phase-resolved spectra.  

{\bf Consideration of the alternative BB+BB model:} 
Although the BB+BB model's fit quality is formally acceptable (Table 1),  there are a number of  serious  deficiencies: 
\begin{itemize}\vspace{-0.5cm}
\item The  $\chi^2$ values are worse than those for the BB$\times$GABS model (Table 1). To check the chance occurrence probability for the $\chi^2$ reduction, we have carried out Monte-Carlo simulations (see below)
   following the recommendations of \cite{2002ApJ...571..545P}. The observed  $\chi^2$ improvement  is significant at 99.4\% confidence even without taking into account the evidence for a second feature (below 0.3 keV)  and a similar dependence of the strengths of the two spectral features on the pulsar phase (Figure 3, left panels).
\item Unlike the BB$\times$GABS model, systematic residuals  remain  both in 0.5--0.8 keV  and below 0.3  keV (Fig.\ S1) despite the larger number of model parameters that are allowed to vary (12 in BB+BB vs.\ 11 in BB$\times$GABS; cf.\ middle and bottom sections of Table 1);
\item the size of the emitting area  of the cooler  BB component fitted to the ``peak'' spectrum ($R=102^{+450}_{-70}$ km)  substantially exceeds a conventional NS radius (Fig.\ S1, bottom panel) while fitting with a hydrogen atmosphere model would increase the apparent radius even further\cite{1995lns..conf...71P};
\item the ``dip'' spectrum cannot be fit satisfactorily by a BB+BB model. Even if we allow all parameters  to vary, we find $\chi_{\nu}^2=1.5$ for 22 d.o.f, and the absorbing column becomes unreasonably large ( $N_H=20_{-6}^{+12}\times10^{20}$\,cm$^{-2}$) and barely compatible with that from the phase-integrated fit.  The intrinsically broad BB simply fails to adequately describe the narrow spectral  peak (centered at about 0.64 keV) in the ``dip'' spectrum (Fig.\ S1, top panel).
\end{itemize}

{\bf Statistical comparison of   BB+BB and BB$\times$GABS models:} We calculated the ratio, $r_{\rm obs}=LR_{\rm obs}^{BB\times GABS}/LR_{\rm obs}^{BB+BB}$,  of the likelihood ratios for the  BB$\times$GABS and  BB+BB fits to the actual data. We then simulated 1000 ``fake'' spectra (using XSPEC ``fakeit'' command)  based on the best-fit BB+BB model (null-hypothesis). These spectra mimic the expected noise fluctuations in a given exposure time (which was the same as for the real data). The same response and background files have been used.
  Each of the simulated spectra was then re-fitted with the BB+BB and BB$\times$GABS model with the temperatures, radii and the line  parameters allowed to vary. For each of the fake spectra  ($i=1,...,1000$) we also recorded the ratio, $r_i=LR_i^{BB\times GABS}/LR_i^{BB+BB}$, of the likelihoods for the BB$\times$GABS and  BB+BB fits. We then  calculated the number of instances with $r_i >r_{\rm obs}$ and found only six such cases.  Therefore, for the real data the BB$\times$GABS model is statistically  preferred  over the BB+BB at a 99.4\% confidence level even without taking into account the evidence for a second spectral feature with a factor of two lower energy and  progressive strengthening of the features from the pulse maximum to the pulse minimum. 

\bibliography{1221378Bibliography}

\begin{thebibliography}{10}
\bibitem{note3}
A simpler model, BB+PL, does not produce a satisfactory fit.
\bibitem{note2}
This $N_{\rm H}$ value is used in all the subsequent phase-resolved spectral
  fits because having $N_{\rm H}$ tied and fitting for it in the phase-resolved
  spectra results in a similar but more uncertain value.
\end{thebibliography}

\bibliographystyle{Science}




\end{document}